\documentclass[]{spie}  
\usepackage[]{graphicx}
\usepackage[]{subfigure} 
\usepackage{epsfig}
\usepackage{amsmath}
\usepackage{tabularx}
\usepackage{dcolumn}
\usepackage{nicefrac}
\newcolumntype{d}[1]{D{.}{.}{#1}}

\title{Optimal control of charge transfer} 

\author{Jan Werschnik\supit{a} and E.K.U. Gross\supit{a}
\skiplinehalf
\supit{a}Freie Universit{\"a}t, Arnimallee 14, Berlin, Germany;
}

\authorinfo{Further author information: (Send correspondence to E.K.U.G.)\\J.W.: E-mail: jan.werschnik@physik.fu-berlin.de\\  E.K.U.G.: E-mail: hardy@physik.fu-berlin.de}

\def\ra{{\,\rangle }}

\def\infintd3r{ \int_{-\infty}^\infty d^3r\,}
\def\intd3r{ \int d^3r\,}

\def\laplace1d{\frac{d^2}{dx^2}}
\def\plaplace1d{\frac{d^2}{d{x'}^2}}

\def\padr2{\frac{\partial^2}{\partial r^2}}

\def\bei{\begin{itemize}}
\def\eei{\end{itemize}}
\def\bee{\begin{enumerate}}
\def\eee{\end{enumerate}}

\def\TDSE{time-dependent Schr\"odinger equation}

\def\octopus{OCTOPUS}
\newfont{\tensy}{cmsy10}
\newcommand{\im}[1]{\,\mathrm{Im}#1}

\newcommand{\chem}[1]{{$\fontdimen16\tensy=3.0pt\fontdimen17\tensy=3.0pt \mathrm{#1}$}}

\def\bepsilon{\boldsymbol \epsilon}  
\def\bmu{\boldsymbol \mu} 

\newcommand{\eref}[1]{Equation~(\ref{#1})}

\newcommand{\sref}[1]{Sect.~\ref{#1}}
\newcommand{\fref}[1]{Fig.~\ref{#1}}
\newcommand{\tref}[1]{Table~\ref{#1}}

\newcommand{\Fref}[1]{Fig.\ \ref{#1}}

\newcommand{\cref}[1]{Chap.\ \ref{#1}}

\newcommand{\opencircle}{\mbox{\Large$\circ\,$}}  
\newcommand{\opensquare}{\mbox{$\rlap{$\sqcap$}\sqcup$}}

\newcommand{\dotted}{\protect\mbox{${\mathinner{\cdotp\cdotp\cdotp\cdotp\cdotp\cdotp}}$}}
\newcommand{\dashed}{\protect\mbox{-\; -\; -\; -}}
\newcommand{\broken}{\protect\mbox{-- -- --}}

\newcommand{\chain}{\protect\mbox{--- $\cdot$ ---}}
\newcommand{\dashddot}{\protect\mbox{--- $\cdot$ $\cdot$ ---}}
\newcommand{\full}{\protect\mbox{------}}

\def\bea{\begin{eqnarray}}
\def\eea{\end{eqnarray}}
\def\ben{\begin{equation}}
\def\een{\end{equation}}

\def\sss{\scriptscriptstyle\rm}

\def\1var{(\bx_1...\bx\N)}

\def\br{{\bf r}}

\def\bx{{\bf x}}

\def\N{_{\sss N}}

\def\sph_int{ {\int d^3 r}}

\begin{document} 
  \maketitle 

\begin{abstract}
In this work, we investigate how and to which extent a quantum system
can be driven along a prescribed path in space by a suitably tailored laser pulse. The laser field is calculated with the help of quantum optimal control theory employing a time-dependent formulation for the control target. 
Within a two-dimensional (2D) model system we have successfully optimized laser fields for two distinct charge transfer processes. The resulting laser fields can be understood as a complicated interplay of different excitation and de-excitation processes in the quantum system.

\end{abstract}



\section{INTRODUCTION}
\label{sec:intro}  
Since the first realization of the laser by T.H. Maiman in 1960
physicists and chemists dream about coherently controlling quantum systems using laser fields. For example, laser pulses may be applied to create and break a particular bond in a molecule, to control charge transfer within molecules, or to optimize high harmonic generation.
%
The first approach to break a certain bond in a molecule using a laser tuned on resonance and initiating a resonance catastrophe failed. The molecule was distributing the energy internally too quickly, so that the specific bond did not break but instead the whole molecule was ``heated''. To overcome this so-called internal vibrational relaxation (IVR) a smarter excitation strategy and further technological improvements were necessary.

With the advent of femtosecond laser pulses in the 1980's and the sophisticated technology\cite{WLPW92} for shaping  the laser pulses the goal of controlling a complex chemical reaction with coherent light was finally achieved: For example, in 1998 Assion {\it et al.}\ \cite{A98} showed that the product ratio \chem{CpFeCOCl^+}/\chem{FeCl^+} of the organo-metallic compound (\chem{CpFe(CO)_2Cl}) can be either maximized or minimized by a specially tailored light pulse; or in 2001, Levis {\it et al.}\ \cite{LMR2001} demonstrated a rearrangement of molecular fragments. 
In both of these experiments adaptive laser pulse shaping techniques\cite{WLPW92,JR92} have been applied, i.e., a computer analyzes the outcome of the experiment and modifies the laser pulse shape to optimize the yield of a predefined reaction product. This process is repeated until the optimal laser pulse is found. The number of experiments based on this so-called closed-learning-loop (CLL) is growing constantly, see for example Refs.~\citenum{B97,B2000,BDNG2001,HWCZM2002,BDKG2003,D2003,PWWSG2005}.
Recently, the pulse shaping techniques have been extended to allow also for polarization shaping\cite{VKNNG2005, POS2006, PWWL2006a, PWWL2006b}, i.e., experimentalists can indepentendly shape polarization, amplitude, and phase.

Similar progress in quantum control can also be observed on the theoretical side starting in the 1980's with rigouros statements about controllability\cite{HTC83,PDR88,K89,RSDRP95,SFS2001}, faster algorithms to calculate optimal laser pulses\cite{TKR86, TR86, TKO92,ZBR98, PK2002,MT2003,WG2005}, and a growing number of investigations for different types of systems and applications\cite{AKT93,AKT97, RVMK2000, GHV2003,D2003,B2004,KPKS2004}. A large part of these studies employ a mathematical technique called optimal control theory (OCT)\cite{SW97, V2000,B2005} well known from engineering. The employed numerical schemes have been designed to reach a predefined target at the end of a finite time-interval. 
Recently, these schemes have been extended to deal with time-dependent targets\cite{OTR2004}, meaning that we can also control the path the quantum system takes to the desired target. Little is known about the controllability\cite{KM2004, KM2005, mySWG2005, janphd} of such objectives. 

In this article we employ the time-dependent target formalism to optimize laser fields that drive a charge along two distinct routes in 2D quadruple well. We allow the laser to have two independent components which would require full polarization shaping in the  experimental realization. The article is organized in the following way: In \sref{sec:theory} we explain the basics of quantum optimal control theory for time-dependent targets and the algorithm which we use to optimize the laser field.  The developed technique is then applied to a model system which is discussed in \sref{sec:model}. %
The control objectives, i.e., the different pathways and the results of the optimizations are shown in \sref{sec:results}. We conclude the article with a summary and a short outlook in \sref{sec:conclusion}.


%

\section{THEORY AND ALGORITHM} 
\label{sec:theory}
%
%
Let us consider an electron in an external potential $V({\bf r})$ under the influence of a laser field. The laser pulse is assumed to propagate in $z$~direction and therefore has two polarizations, $x$ and $y$, so that a test charge is accelerated only in the plane perpendicular to the $z$~axis. Given an initial state $\Psi({ \bf r },0)=\phi({ \bf r })$ the time evolution of the electron is described by the  time-dependent Schr\"odinger equation with the laser field modeled in dipole approximation (length gauge)
\begin{eqnarray}
i\frac{\partial}{\partial t}\Psi({ \bf r },t)&=&\widehat{H}\Psi({ \bf r },t),\label{1SE}\\
\widehat{H}&=&\widehat{H}_0-\hat{{\boldsymbol \mu}}{\boldsymbol \epsilon}(t),\\
\widehat{H}_0&=&\widehat{T}+\widehat{V},
\end{eqnarray}
(atomic units [a.u.] are used throughout: $\hbar = m =e = 4\pi \epsilon_0 = 1 $).  Here, $\hat{{\boldsymbol \mu}}=(\hat{\mu}_x,\hat{\mu}_y)  $ is the dipole operator, and ${\boldsymbol \epsilon}(t)=(\epsilon_x(t),\epsilon_y(t))$ is the time-dependent electric field. For an electron we simply have $\hat{\bmu} = -\hat{\br}$. The kinetic energy operator is $\widehat{T}=-\frac{\nabla^2}{2}$.
\subsection{Derivation of control equations}
Our goal is to control the time evolution of the electron by the external field in a way that the time averaged expectation value of the target operator $\widehat{O}(t)$ is maximized. Mathematically,  
the optimization goal corresponds to the maximization of  the functional
\begin{eqnarray} \label{eq:J1}
J_1[\Psi]&=&\frac{1}{T} \int_0^T \!\! dt \,\, \langle \Psi(t)|\widehat{O}(t)|\Psi(t)\rangle,
\end{eqnarray} 
where $\widehat{O}(t)$ is assumed to be positive-semidefinite and defined by
\begin{eqnarray}
\widehat{O}(t)&=&\widehat{O}_{1}(t)+ 2 T \delta(t-T) \, \widehat{O}_{2},
\end{eqnarray} 
so we can also include targets in our formulation that only depend on the final time $T$\cite{ZBR98,ZR98,K89}. A few examples will be discussed at the end of this section.
%
%

The functional $J_1[ \Psi]$ will be maximized subject to a number of physical constraints.
The idea is to cast also these constraints in a suitable functional form and then calculate the total variation. Subsequently, we set the total variation to zero and find a set of coupled partial differential equations \cite{K89,PDR88}. The solution of these equations will yield the desired laser field ${ \boldsymbol \epsilon }(t)$.

In more detail: Optimizing $J_1$ may possibly lead to fields with very high, or even infinite total intensity. In order to avoid these strong fields, we include an additional term in the functional which penalizes the total energy of the field.
\begin{eqnarray}
J_2[{ \boldsymbol \epsilon }] &=& - \alpha\int_0^T \!\! dt \,\, { \boldsymbol \epsilon }^2(t). 
\end{eqnarray} 
The penalty factor $\alpha$ is a positive parameter used to weight this part of the functional against the other parts.
The constraint that the electron's wave-function has to fulfill the time-dependent Schr\"odinger equation is expressed by
\begin{eqnarray}
  J_3[{ \boldsymbol \epsilon },\Psi,\chi]&=&
    - 2 \im \int_{0}^{T}\!\!  dt \,\, \left\langle\chi(t) \left| \left(i\partial_t
    -\widehat{H}\right) \right| \Psi(t)\right\rangle.
\end{eqnarray} 
with a Lagrange multiplier $\chi({ \bf r },t)$. $\Psi({ \bf r },t)$ is the wave-function driven by the laser field ${ \boldsymbol \epsilon }(t)$.
The Lagrange functional has the form
\begin{eqnarray}
J[\chi,\Psi,{ \boldsymbol \epsilon }] = J_1[\Psi] + J_2[{ \boldsymbol \epsilon }] + J_3[\chi,\Psi,{ \boldsymbol \epsilon }].
\end{eqnarray} 
Setting the variations of the functional with respect to $\chi$, $\Psi$, and ${ \boldsymbol \epsilon }$ independently to zero yields
\begin{eqnarray}
 \alpha \epsilon_j(t) &=& -\im\langle\chi(t)|\hat{\mu}_j|\Psi(t)\rangle, \label{eq:field} \qquad j=x,y,z\\
0 &=& \left( i \partial_t - \widehat{H} \right) \Psi({ \bf r },t), \label{eq:SE}\\
\Psi({ \bf r },0) &=& \phi({ \bf r }),\label{eq:SE_init}\\
 &&\left(i\partial_t - \widehat{H}\right)\chi({ \bf r },t) + \frac{i}{T}\widehat{O}_{1}(t) \Psi({ \bf r },t)=\nonumber\\
\label{eq:INHSE}
&&i\left(\chi({ \bf r },t)-\widehat{O}_{2}(t) \Psi({ \bf r },t)\right)\delta(t-T).
\end{eqnarray}
Equation (\ref{eq:field}) determines the field from the wave-function $\Psi({ \bf r },t)$ and the Lagrange multiplier $\chi({ \bf r },t)$. While
Equation (\ref{eq:SE}) is a time-dependent Schr\"odinger equation for $\Psi({ \bf r },t)$ starting from a given initial state $\phi({ \bf r })$ and driven by the field ${ \boldsymbol \epsilon }(t)$. 
If we require the Lagrange multiplier $\chi({ \bf r },t)$ to be continuous, we can solve the following two equations\cite{mySWG2005} instead of Equation (\ref{eq:INHSE}):
\begin{eqnarray}
\label{eq:INHSE2}
 \left(i\partial_t - \widehat{H}\right)\chi({ \bf r },t)&=&-\frac{i}{T}\widehat{O}_{1}(t)\Psi({ \bf r },t),\\
\label{eq:INHSE3}
\chi({ \bf r },T) &=&\widehat{O}_{2} \Psi({ \bf r },T).
\end{eqnarray}
%
%
%
%
%
%
%
%
%
Hence, the Lagrange multiplier satisfies an inhomogeneous Schr\"odinger equation with an initial condition at $t=T$. Its solution can be formally written as
\begin{eqnarray}
\label{eq:SOL_INHSE}
\chi({ \bf r },t)&=&\widehat{U}_{t_0}^{t}\chi({ \bf r },t_0)-
       \frac{1}{T}\int_{t_0}^{t} \!\! d\tau \,\, \widehat{U}_{\tau}^{t}\left(\widehat{O}_{1}(\tau) \, \Psi({ \bf r },\tau)\right),
\end{eqnarray}
where $\widehat{U}_{t_0}^{t}$ is the time-evolution operator defined as  $\widehat{U}_{t_0}^{t}=\mathcal{T}\exp\left(-i \int_{t_0}^{t} \!\! dt' \,\, \widehat{H}(t')\right)$.\\
The set of equations that we need to solve is now complete:  (\ref{eq:field}), (\ref{eq:SE}),  (\ref{eq:SE_init}), (\ref{eq:INHSE2}) and (\ref{eq:INHSE3}). To find an optimal field ${ \boldsymbol \epsilon }(t)$ from these equations we use an iterative algorithm which is discussed in the next section.
%
We conclude the derivation with a few examples for the target operator $\widehat{O}(t)$.
\paragraph{Final-time control:}
Since the approach is a generalization of the traditional optimal control formulation given in \cite{K89,ZBR98,ZR98} we first observe that the latter is trivially recovered as a limiting case by setting
\begin{eqnarray}
\widehat{O}_1(t) = 0, \qquad \widehat{O}_2 = \widehat{P} = | \Phi_f \rangle \langle \Phi_f|.
\end{eqnarray}
Here $|\Phi_f\ra$ represents the target final state which the propagated wave-function $\Psi({ \bf r },t)$ is supposed to reach at time $T$. 
In this case the target functional reduces to\cite{K89,ZBR98}
\begin{eqnarray}
J_1 = \langle \Psi(T) | \widehat{P} | \Psi(T) \rangle = | \langle \Psi(T) | \Phi_f \rangle| ^2 .
\end{eqnarray}
The target operator may also be local in space\cite{ZR98}. If we choose $\widehat{O}_1(t) = 0$ and $\widehat{O}_2 = \delta({ \bf r }-{ \bf r }_0)$ (the density operator), we can maximize the probability density in ${ \bf r }_0$ at $t=T$.
\begin{eqnarray}
\label{eq:loc_op}
J_1 = \int \!\! d{ \bf r } \,\, \langle \Psi(T)| \widehat{O}_2 |\Psi(T) \rangle = n({ \bf r }_0,T). 
\end{eqnarray}
%
%
\paragraph{Wave-function-follower:}
The most ambitious goal is to find the pulse that forces the system to follow a predefined wave-function $\Phi({ \bf r },t)$. If we choose
\begin{eqnarray}
\widehat{O}_{1}(t) &=& |\Phi(t)\rangle\langle\Phi(t)|, \\
\widehat{O}_{2}&=&0,
\end{eqnarray}
the maximization of the time-averaged expectation value of $\widehat{O}_{1}(t)$ with respect to the field ${ \boldsymbol \epsilon }(t)$ becomes almost equivalent to the inversion of the Schr\"odinger equation, i.e., for a given function $\Phi({ \bf r },t)$ we find the field $ { \boldsymbol \epsilon }(t)$ 
so that the propagated wave-function $\Psi({ \bf r },t)$ comes as close as possible to the target $\Phi({ \bf r },t)$ in the space of admissible control fields.
%
%
%
%
\paragraph{Moving density:}
In this article we will focus on the following type of target. 
The operator used in Equation~(\ref{eq:loc_op}) can be generalized to
\begin{eqnarray}
\widehat{O}_{1}(t)&=&\delta({ \bf r }-{\bf r }_0(t)),\\
J_1 &=& \frac{1}{T} \int_0^T \!\! dt \,\, \langle \Psi(t) |  \delta({ \bf r }-{ \bf r}_0(t)) | \Psi(t) \rangle  \nonumber\\
\nonumber
&=& \frac{1}{T} \int_0^T \!\! dt \,\, n({\bf r}_0(t),t).
\end{eqnarray}
$J_1$ is maximal if the field is able to maximize the density along the given trajectory ${\bf r }_0(t)$. In \sref{sec:results} we will analyze the optimized fields calculated for two different trajectories of our 2D model system.
In the numerical implementation the $\delta$-function is approximated by a narrow Gaussian
\bea
\label{eq:2dtargetop}
\hat{O}(\br) &=& \delta({ \bf r }-{\bf r }_0(t))\\
               &\approx& \frac{1}{\sigma \sqrt{\pi}}\exp\left[-\frac{({ \bf r }-{\bf r }_0(t))^2}{2 \sigma^2}\right].
\eea
%
\subsection{Algorithm for time-dependent control targets}
\label{sec:tdctrlalg}
%
Equipped with the control equations (\ref{eq:field}), (\ref{eq:SE}), (\ref{eq:SE_init}),  (\ref{eq:INHSE2}), and (\ref{eq:INHSE3}), we have to find an algorithm to solve these equations for ${\boldsymbol \epsilon}(t)$.
In the following we describe the scheme discussed in Refs.~\citenum{OTR2004, mySWG2005}.
%
 A monotonic convergence in $J$ can be proven if $\eta \in [0,1]$ and $\xi \in [0,2]$ \cite{MT2003}. 
%

The algorithm starts with propagating $\Psi^{(0)}(0)=\phi_i$ forward in time with an initial guess for the laser field $\bepsilon(t)^{(0)}$,
\begin{equation}
\nonumber
\begin{array}{l c c l c c c c }
  {\mbox{step 0:}} \,\, & \Psi^{(0)}(0) & \overset{{\bepsilon}^{(0)}(t)}{\longrightarrow} &
 \Psi^{(0)}(T). &  &  &  &
\end{array}
\end{equation}
The backward propagation of $\chi^{(0)}(t)$ is started from $\chi^{(0)}(T)=0$ solving an inhomogeneous \TDSE~which requires $\Psi^{(0)}(t)$ as input (with $k=0$),
\begin{equation}
\label{eq:tdctrlstepk}
\begin{array}{l c c l c c c c }
 {\mbox{step k:}} \,\, & & & \left[ \Psi^{(k)}(T) \right. & \overset{{\boldsymbol \epsilon}^{(k)}(t)}{\longrightarrow} &
 \left. \Psi^{(k)}(0) \right] & & \\
                   & & & \chi^{(k)}(T) & \overset{\widetilde{{\boldsymbol \epsilon}}^{(k)}(t)
,\:\:\Psi^{(k)}(t)\:\: }{\longrightarrow} & \chi^{(k)}(0)\, . & &
\end{array}
\end{equation}
The brackets indicate that the storage of the wave function $\Psi^{(0)}(t)$ can be avoided if we propagate it backwards in time as well using $\bepsilon^{(0)}(t)$. The backward propagation of $\chi^{(0)}(t)$ requires the laser field determined by ($k=0$),
\begin{eqnarray}
\label{feld1}
\widetilde{\epsilon}_j^{(k)}(t) &=& (1-\eta)\epsilon_j^{(k)}(t) 
-  \frac{\eta}{\alpha}\im\, \langle\chi^{(k)}(t)|\hat{\mu}_j|\Psi^{(k)}(t)\rangle,\,\,\ j=x,y . \,\,\,\,\,\,\,\,\,
\end{eqnarray}
The next step is to start a forward propagation of  $\Psi^{(1)}(0) = \phi_i$ ($k=0$)
\begin{equation}
\nonumber
\begin{array}{l c c l c c c c }
                   & & & & & \left[ \chi^{(k)}(0) \right. & \overset{\widetilde{{\boldsymbol \epsilon}}^{(k)}(t) 
 ,\:\:\Psi^{(k)}(t)\:\: }{\longrightarrow} & \left. \chi^{(k)}(T) \right] \\
                   & & & & & \left[ \Psi^{(k)}(0) \right. & \overset{{\boldsymbol \epsilon}^{(k)}(t)}
{\longrightarrow} & \left. \Psi^{(k)}(T) \right] \\
                   & & & & & \Psi^{(k+1)}(0) & \overset{{\boldsymbol \epsilon}^{(k+1)}(t)}
{\longrightarrow} & \Psi^{(k+1)}(T) .\\
\end{array}
\end{equation}
and calculate the laser field 
\begin{eqnarray}
\label{feld2}
\epsilon_j^{(k+1)}(t) &=& (1-\xi)\widetilde{\epsilon}_j^{(k)}(t)
- \frac{\xi}{\alpha}\im\, \langle\chi^{(k)}(t)|\hat{\mu}_j|\Psi^{(k+1)}(t)\rangle, \,\,\ j=x,y.\,\,\,\,\,\,\,\,\, 
\end{eqnarray}
%
%
After the time evolution is complete we can close the loop and continue with \eref{eq:tdctrlstepk}.
The whole scheme can be depicted symbolically by:
\begin{equation}
\label{eq:tdctrlscheme}
\begin{array}{l c c l c c c c }
  {\mbox{step 0:}} \,\, & \Psi^{(0)}(0) & \overset{{\bepsilon}^{(0)}(t)}{\longrightarrow} &
 \Psi^{(0)}(T) &  &  &  &\\
 {\mbox{step k:}} \,\, & & & \left[ \Psi^{(k)}(T) \right. & \overset{{\boldsymbol \epsilon}^{(k)}(t)}{\longrightarrow} &
 \left. \Psi^{(k)}(0) \right] & & \\
                   & & & \chi^{(k)}(T) & \overset{\widetilde{{\boldsymbol \epsilon}}^{(k)}(t) 
,\:\:\Psi^{(k)}(t)\:\: }{\longrightarrow} & \chi^{(k)}(0) & &\\
                   & & & & & \left[ \chi^{(k)}(0) \right. & \overset{\widetilde{{\boldsymbol \epsilon}}^{(k)}(t) 
 ,\:\:\Psi^{(k)}(t)\:\: }{\longrightarrow} & \left. \chi^{(k)}(T) \right] \\
                   & & & & & \left[ \Psi^{(k)}(0) \right. & \overset{{\boldsymbol \epsilon}^{(k)}(t)}
{\longrightarrow} & \left. \Psi^{(k)}(T) \right] \\
                   & & & & & \Psi^{(k+1)}(0) & \overset{{\boldsymbol \epsilon}^{(k+1)}(t)}
{\longrightarrow} & \Psi^{(k+1)}(T) .\\
\end{array}
\end{equation}
%
%
%
%
%
%
%
%
%
%
%
%
%
%


The choice of $\eta$ and $\xi$ completes the algorithm. $\xi = 1$ and $\eta = 1$ correspond to the algorithm suggested in Ref.~\citenum{ZR98}, while the choice $\xi = 1$ and $\eta = 0$ is analogous to the method used in Ref.~\citenum{K89} with a direct feedback of $\Psi^{(k)}(t)$. Further choices are discussed in Ref.~\citenum{MT2003}. A more detailed discussion on the convergence of this algorithm with exclusively time-dependent target operators and a modified version of the target functional can be found in Refs.~\citenum{ioanadipl,janphd}.
An iteration scheme which also incorporates spectral restrictions\cite{WG2005} on the laser pulse is presented and analyzed in Ref.~\citenum{janphd}. 

Following the algorithm described above, one needs 5 propagations per iteration-step (avoiding the storage of the wave-function). Within the 2nd order split-operator scheme each time step requires 4 Fast Fourier Transforms (FFT) \cite{FFTW98}, because we have to know the wave function and the Lagrange multiplier in real space at every time-step to be able to evaluate the field from Eq.  (\ref{eq:field}). This sums up to $2*10^6$ FFTs per $10^5$ time steps and iteration. 


\section{MODEL AND COMPUTATIONAL DETAILS} 
\label{sec:model}
\begin{figure}[tpb]
\centering 
\subfigure[]{
  \label{fig:adw2Ds1} 
  \resizebox{0.46\textwidth}{.22\textheight}{\includegraphics{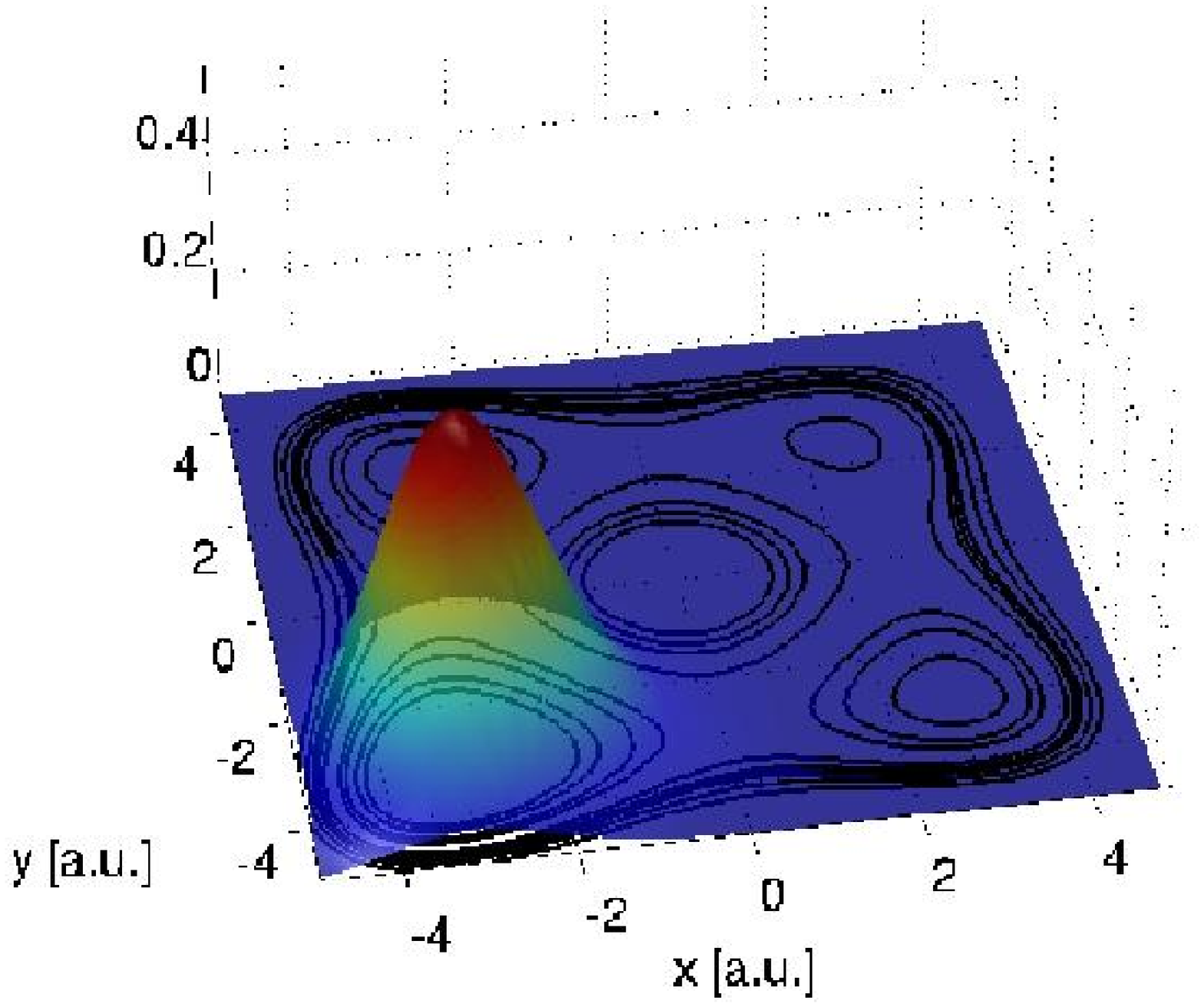}}

    }
\subfigure[]{
  \label{fig:adw2Ds2} 
\resizebox{0.46\textwidth}{.22\textheight}{\includegraphics{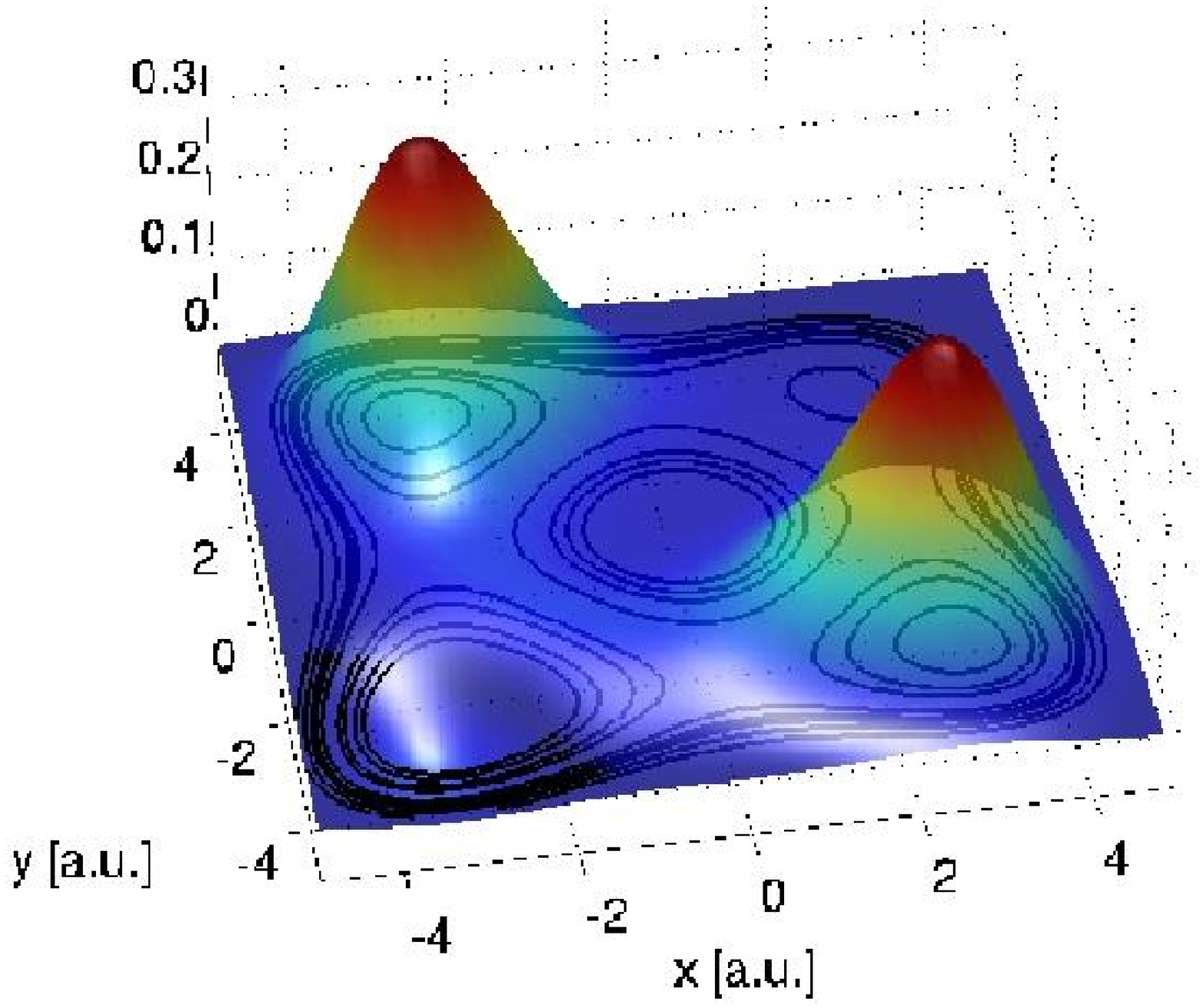}}

}\\
\subfigure[]{
  \label{fig:adw2Ds3}
  \resizebox{0.46\textwidth}{.22\textheight}{\includegraphics{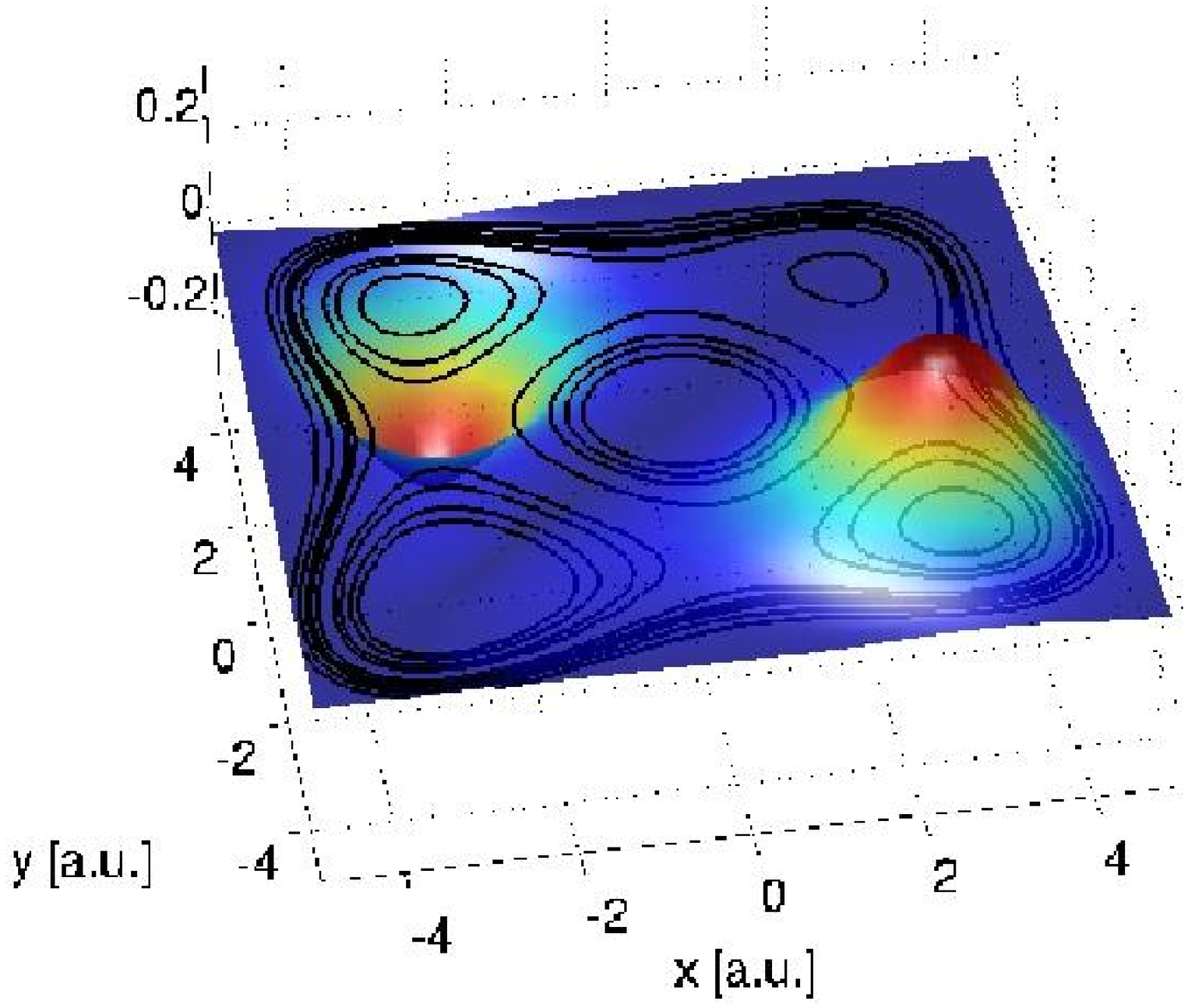}}
}
\subfigure[]{
  \label{fig:adw2Ds4} 
  \resizebox{0.46\textwidth}{.22\textheight}{\includegraphics{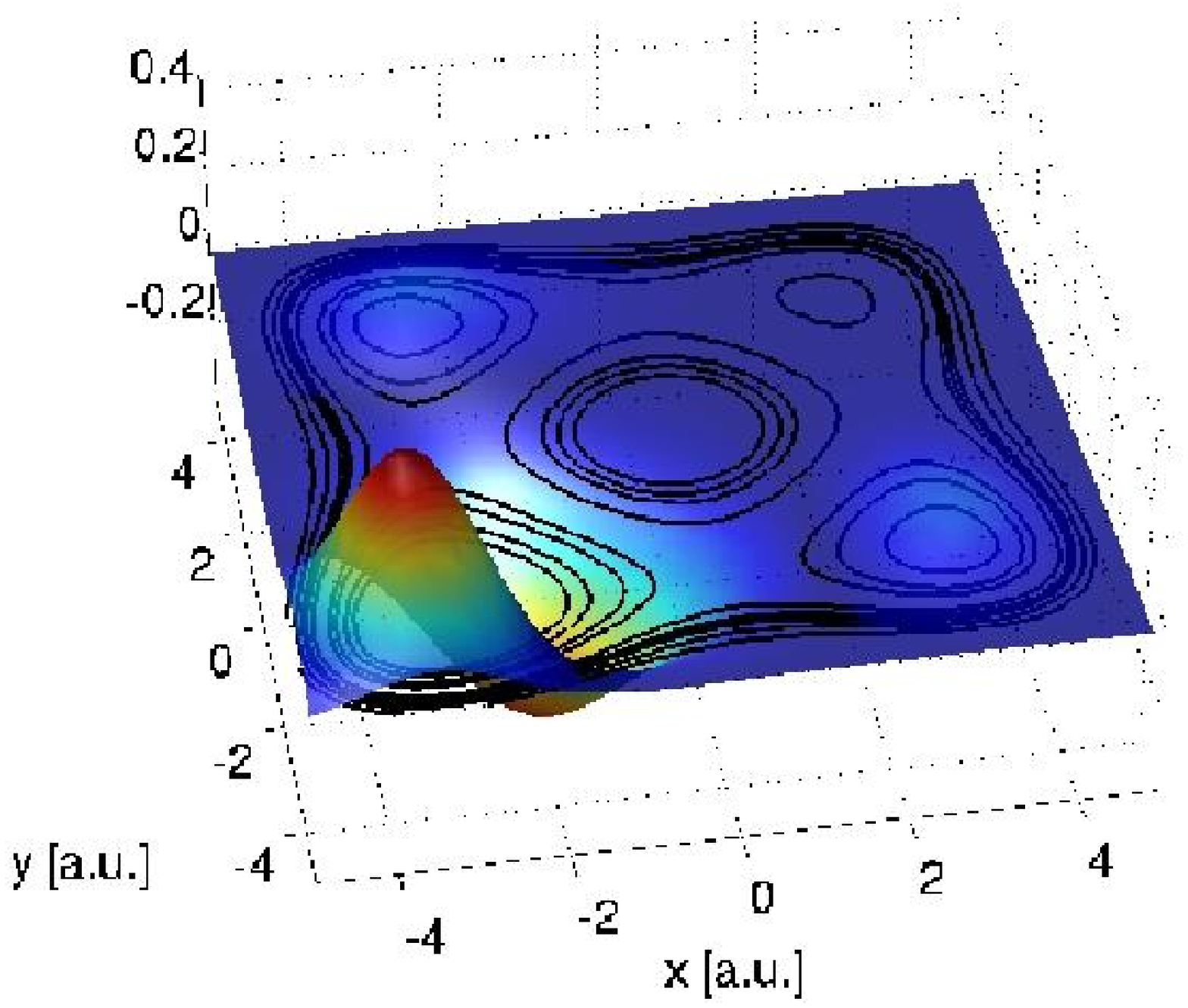}}

    }\\
\subfigure[]{
  \label{fig:adw2Ds5}
  \resizebox{0.46\textwidth}{.22\textheight}{\includegraphics{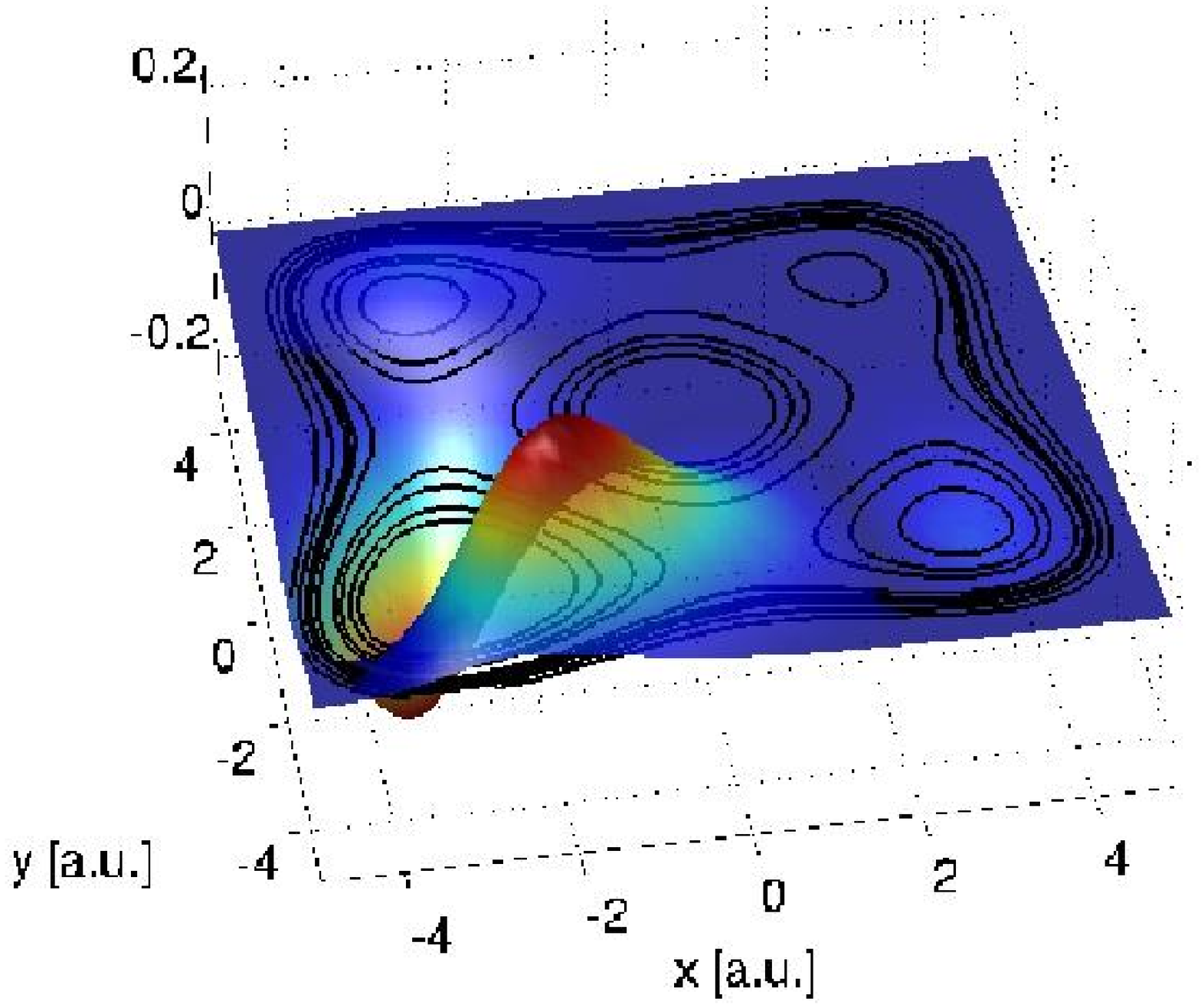}}

}
\subfigure[]{
  \label{fig:adw2Ds6}
  \resizebox{0.46\textwidth}{.22\textheight}{\includegraphics{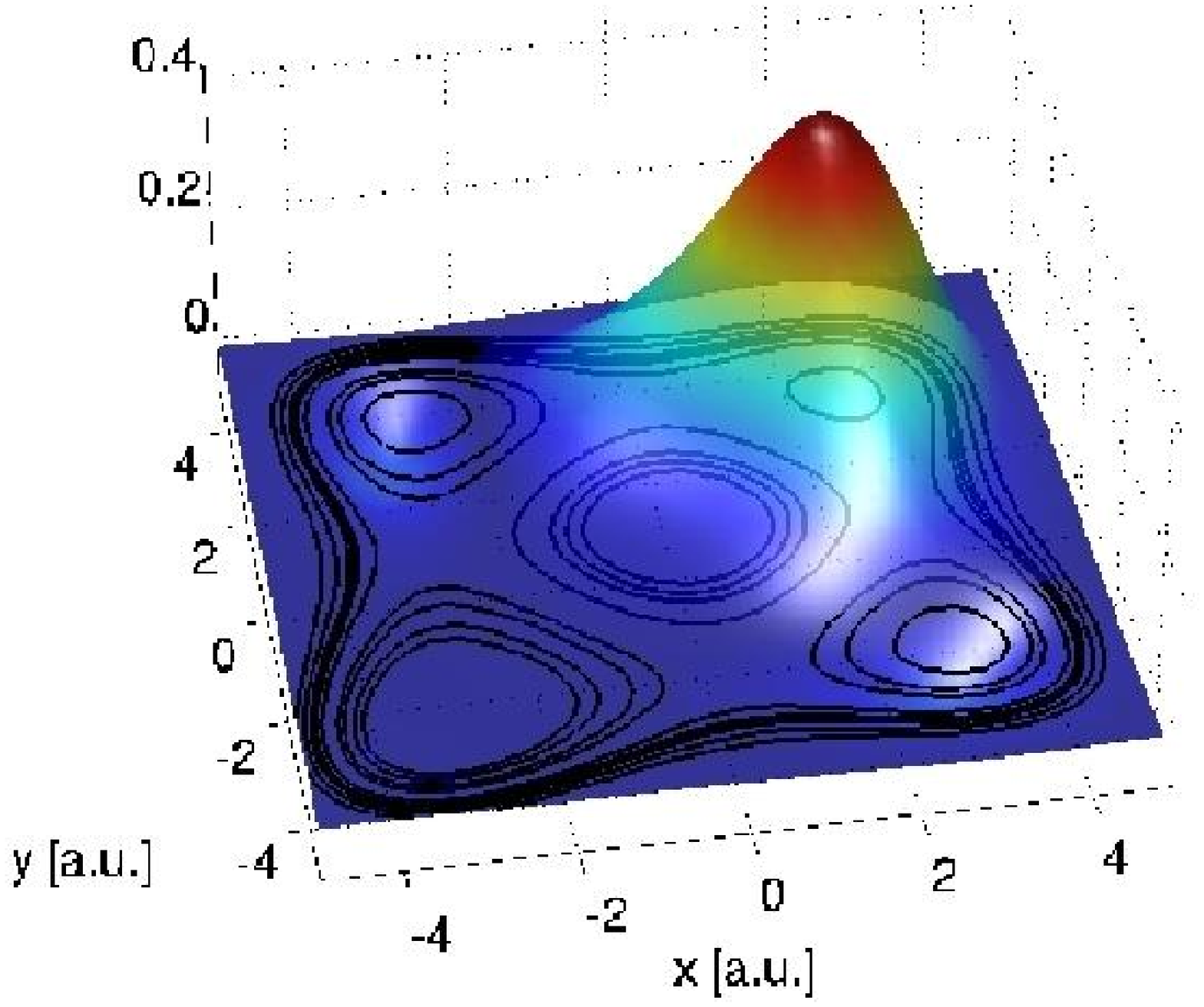}}

}

\caption[Six lowest eigenstates for the 2D asymmetric quadruple well.]{(Color online). The six lowest eigenstates of the asymmetric quadruple well structure: (a) ground state, (b) 1st, (c) 2nd, (d) 3rd, (e) 4th, (f) 5th excited state. The contour lines indicate the potential surface.}
\label{fig:pot2D_states}
\end{figure}

\begin{figure}[!h]
\centering 
\subfigure[]{
  \label{fig:pot2Dindirect}
  \resizebox{0.42\textwidth}{.22\textheight}{\includegraphics{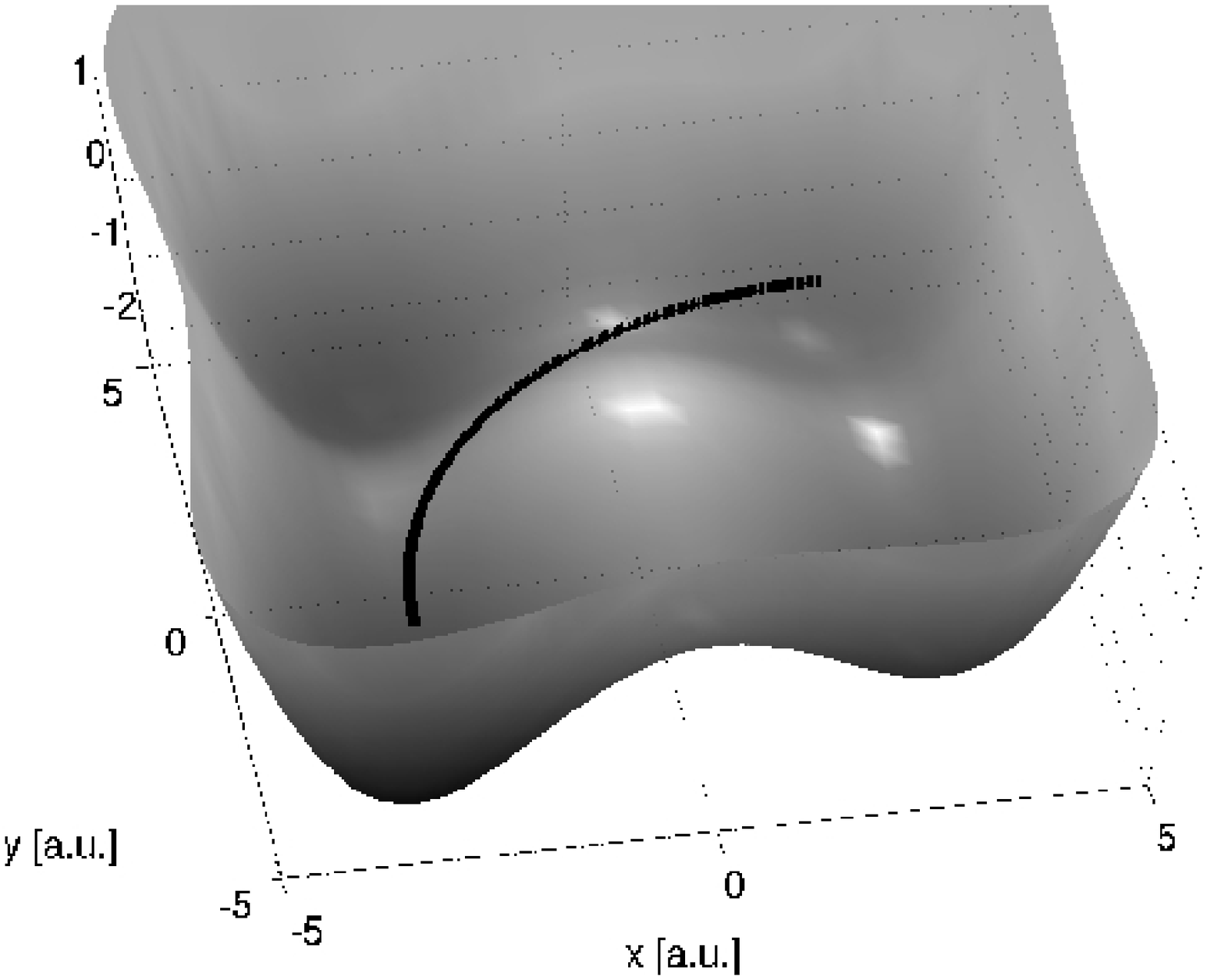}}
}
\subfigure[]{
  \label{fig:pot2Ddirect}
  \resizebox{0.42\textwidth}{.22\textheight}{\includegraphics{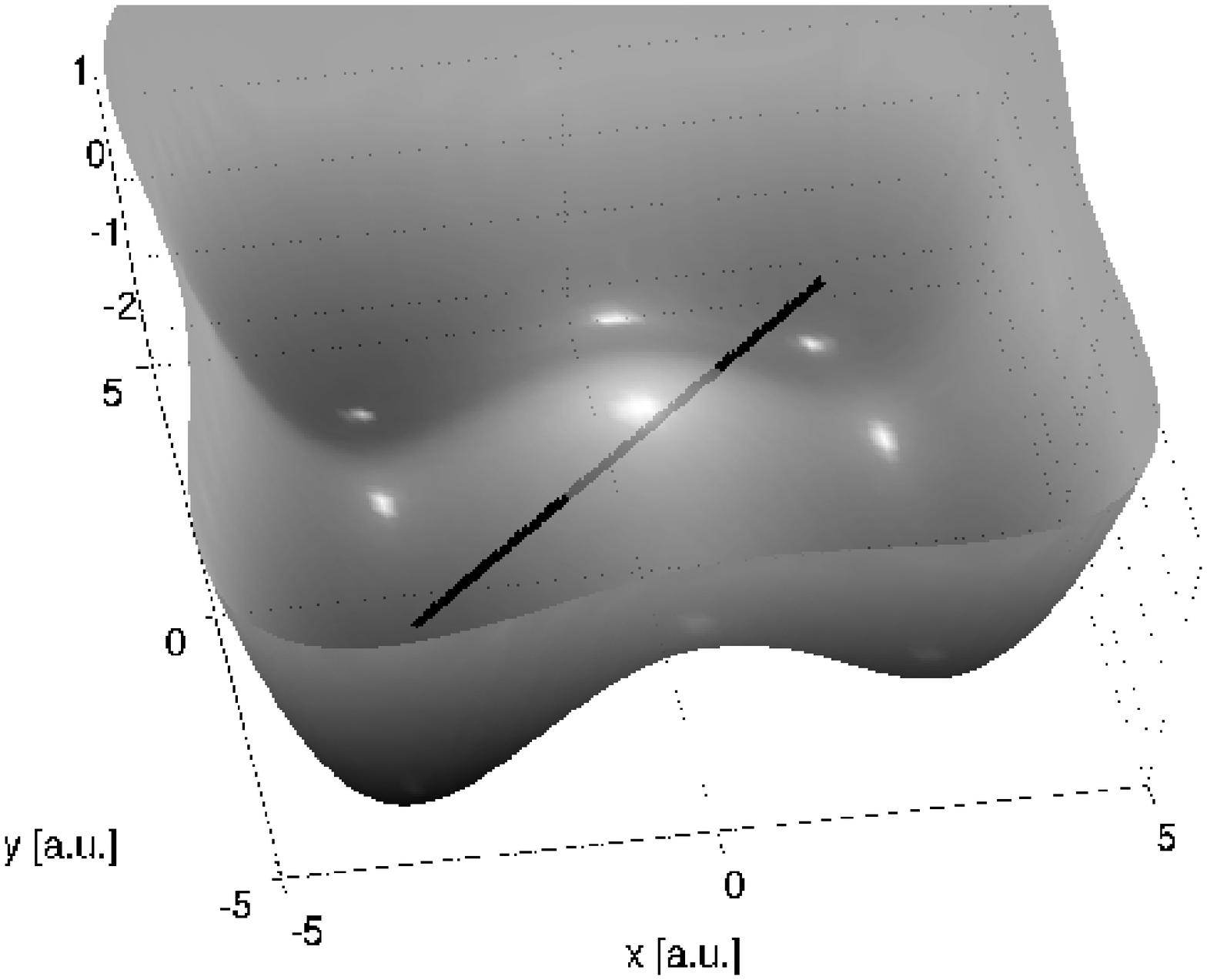}}
 }
\caption[2D Potential and target trajectory.]{The potential surface together with the ``indirect'' (a) and the ``direct'' (b) trajectory. The indirect route goes around the barrier while the direct route leads across the barrier.}
\label{fig:pot2D}
\end{figure}

In this section we define our model system and analyze some of its properties.
The potential surface is given by
\bea
\label{eq:2dpot}
V(x,y) = a (x^4 + y^4) - b (x^2 + y^2) + c(x + y)^3, 
\eea

which describes a structure with a well in each corner of a square and a barrier in the center (see \fref{fig:pot2D}). It is asymmetric with respect to $x=-y$, i.e., the well at $(-3,-3)$ is deeper than the one at $(2.5,2.5)$, but symmetric for $x=y$. The parameters are chosen to be $a=1/64$, $b=1/4$ and  $c=1/256$ for which we show the six lowest eigenstates in \fref{fig:pot2D_states}. In \tref{tab:2Dexc} we have calculated a few low-lying excitation energies necessary for the interpretation of the optimized pulse.

\begin{table}[h!]
\footnotesize
\caption[Excitation energies for the 2D nanostructure.]{\label{tab:2Dexc}Excitation energies in (a.u.) for the 2D asymmetric well, calculated by imaginary time propagation.}
\begin{center}
\begin{tabular}{@{}llllllllll}
 \hline
 \rule[-1ex]{0pt}{3.5ex}               & $|0\ra$ &$|1\ra$   &$|2\ra$  & $|3\ra$& $|4\ra$ & $|5\ra$&$|6\ra$ & $|7\ra$ & $|8\ra$\\
\hline
 \rule[-1ex]{0pt}{3.5ex} $|0\rangle$   & $0.$    &          &         &        &        &        &&&\\
 \rule[-1ex]{0pt}{3.5ex} $|1\rangle$   & $0.7046$  &  $0.$      &         &        &        &        &&&\\
\rule[-1ex]{0pt}{3.5ex}  $|2\rangle$   & $0.7082$  &  $0.0035$  & $0.$      &        &        &        &&&\\
\rule[-1ex]{0pt}{3.5ex}  $|3\rangle$   & $0.8751$  &  $0.1704$  & $0.1669$  & $0.$     &        &        &&&\\
\rule[-1ex]{0pt}{3.5ex}  $|4\rangle$   & $0.9956$  &  $0.2910$  & $0.2875$  & $0.1206$ & $0.$     &        &&&\\
\rule[-1ex]{0pt}{3.5ex}  $|5\rangle$   & $1.1322$  &  $0.4275$  & $0.4240$  & $0.2571$ & $0.1366$ &        &&&\\
\rule[-1ex]{0pt}{3.5ex}  $|6\rangle$   & $1.3875$  &  $0.6828$  & $0.6793$  & $0.5124$ & $0.3918$ & $0.2553$ &&&\\
\rule[-1ex]{0pt}{3.5ex}  $|7\rangle$   & $1.4112$  &  $0.7066$  & $0.7030$  & $0.5362$ & $0.4156$ & $0.2790$ & $0.0238$ &&\\
\rule[-1ex]{0pt}{3.5ex}  $|8\rangle$   & $1.4338$  &  $0.7292$  & $0.7256$  & $0.5587$ & $0.4382$ & $0.3016$ & $0.0463$ & $0.0226$ & \\ 
\rule[-1ex]{0pt}{3.5ex} $|9\rangle$    & $1.4875$  &  $0.7829$  & $0.7793$  & $0.6124$ & $0.4919$ & $0.3553$ & $0.1000$ & $0.0763$ & $0.0537$ \\
\hline
\end{tabular}
\end{center}
\normalsize
\end{table}
%
%
%

%
%


\section{RESULTS} 
\label{sec:results}
Finally, we employ the algorithm presented in \sref{sec:theory} to calculate the laser field that guides the quantum system along the specified path. Both calculations start in the ground state $|\Psi(0)\ra = |\phi_0\ra$ and employ the numerical parameters listed in \tref{tab:tparameter}.
\begin{table}[h!]
\caption[Numerical parameters for 2D calculations.]{\label{tab:tparameter}Numerical parameters for the trajectory optimization of the two-dimensional quantum well structure.}
\begin{center}
\begin{tabular}{@{}lll}
\hline
parameter     & \mbox{indirect/direct} &  \\
\hline
$T$           & $200.0$ & pulse length  \\   
$x_{\mathrm{max}}$  & $10.0$ &  grid size \\
$x_{\mathrm{min}}$  & $-10.0$ & grid size \\
$dx$          & $0.078125$ & grid spacing \\ 
$dt$          & $0.001$  & time step \\
$\alpha_x$    & $0.002$  & penalty factor x-polarization \\
$\alpha_y$    & $0.002$  & penalty factor y-polarization \\
$\epsilon^{(0)}$  & $0.0$ & initial guess \\
$\sigma$    & $0.316$   & width of the target operator \\
\hline
\end{tabular}
\end{center}
\end{table}
%

\subsection{Transfer around the barrier}
The indirect route is described by
\bea
\br^{\mathrm{i}}_0(t) = \left( \begin{matrix} 
  2.5 &-& 5.5 \cos[\pi t/(2\, T)] \cr
  -3.0 &+& 5.5 \sin[\pi t/(2\, T)] \cr
\end{matrix}\right), 
\eea
\begin{figure}[!h]
\centering 
\subfigure[]{
  \label{fig:field2Di_1} 
  \resizebox{0.42\textwidth}{.22\textheight}{\includegraphics*{./2Di_field2.eps}}
    }
\subfigure[]{
  \label{fig:field2Di_2} 
  \resizebox{0.42\textwidth}{.22\textheight}{\includegraphics*{./2Di_fft2.eps}}
}\\
\subfigure[]{
  \label{fig:field2Di_3} 
  \resizebox{0.42\textwidth}{.28\textheight}{\includegraphics*{./2Di_population2.eps}}
}
\subfigure[]{
  \label{fig:field2Di_4} 
  \resizebox{0.42\textwidth}{.28\textheight}{\includegraphics*{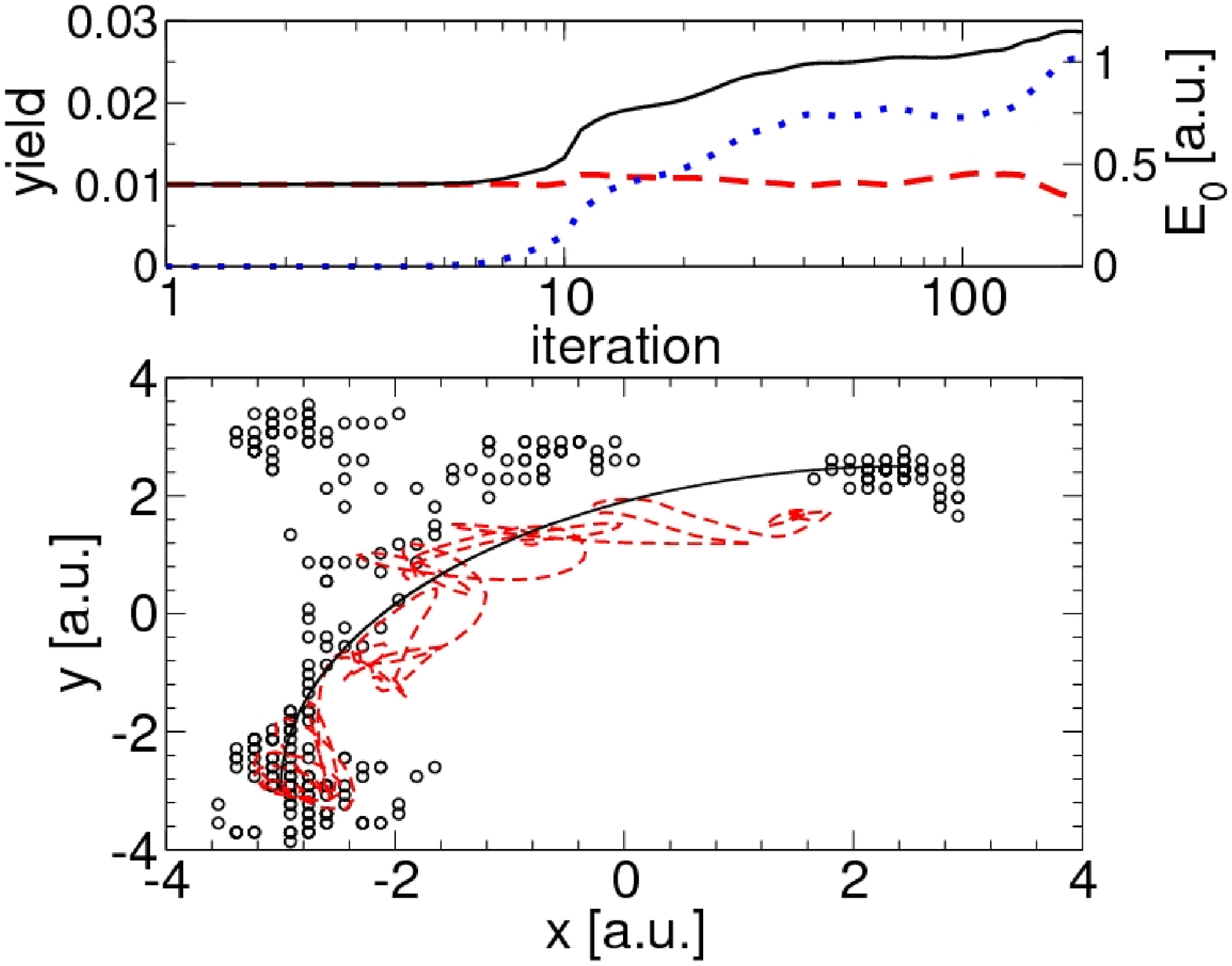}}
}

\caption[Indirect route: Results.]{(Color online). Results for the optimization of the indirect trajectory [see \fref{fig:pot2Dindirect}]. (a) Optimized laser field. (b) Spectrum of optimized laser field. Lower panel: $x$ component. Upper panel: $y$ component. (c) Time evolution of the most important occupation numbers: $|\langle n | \Psi(t) \rangle |^2 $: Top panel: $n=0$ (\full), $n= 4$ (\broken), $n=7$ (\chain). Bottom panel: $n=5$ (\dotted), $n=8$ (\dashddot), $n=9$ (\opensquare). The (\full) line corresponds to the occupation for all states with $n>15$. (d) Lower panel: Target trajectory [(\full) line], the expectation value of the position operator [(\dashed) line], and the position of the density maximum during the propagation (\opencircle). Upper panel: Convergence of the algorithm where the (\dashed) line corresponds to $J$, the (\full) line to $J_1$, and the laser fluence $E_0$ is depicted by the (\dotted) line.}
\label{fig:field2Di}
\end{figure}
%
%
The results of the optimization are presented in \fref{fig:field2Di}. \Fref{fig:field2Di_1} shows that the $x$ (in the upper panel) and the $y$ component of the field (in the lower panel) are quite different in their temporal behavior but similar in strength. The corresponding fluence ($E_0=1.0070$) is rather high.
The spectrum in \fref{fig:field2Di_2} contains significant contributions for  $\omega \in [0;2]$ except for a narrow dip around $\omega=0.5$. This shows that a large number of intermediate states are occupied during the excitation process. 
Looking at the time evolution of the occupation numbers, we see that nearly all states up to the $10$th excited state are significantly involved in the transition process.
In \fref{fig:field2Di_3} we show the occupation numbers with the largest contributions, i.e., the ground state [(\full) line], the $4$th excited [(\dashed) line], the $5$th excited [(\dotted) line], the $7$th excited [(\chain) line], the $8$th excited [(\dashddot) line], and the $9$th excited [(\opensquare) line]. We omit plotting the occupation number of the $3$rd excited state since it is similar to the $4$th. After $t=150$ the pulse shifts nearly all occupation to the $5$th exicted state which is localized at the target position in the right upper well. 

The convergence of the algorithm is shown in the upper panel of \fref{fig:field2Di_4}. We find a fast initial convergence (up to 20 iterations) and a slow improvement of the target functional $J_1$ [(\full) line] for the remaining iterations. 
Although this scheme is in principle monotonically convergent it turns out to be very demanding to achieve an accurate enough solution of the control equations which would then guarantee the monotonicity. To assure the validity of the optimized laser field we have propagated it on a finer grid leading to the same results.

\begin{figure}[htpb]
\centering 
\subfigure[]{
\label{fig:field2Di_6p1}
\resizebox{0.46\textwidth}{.23\textheight}{\includegraphics{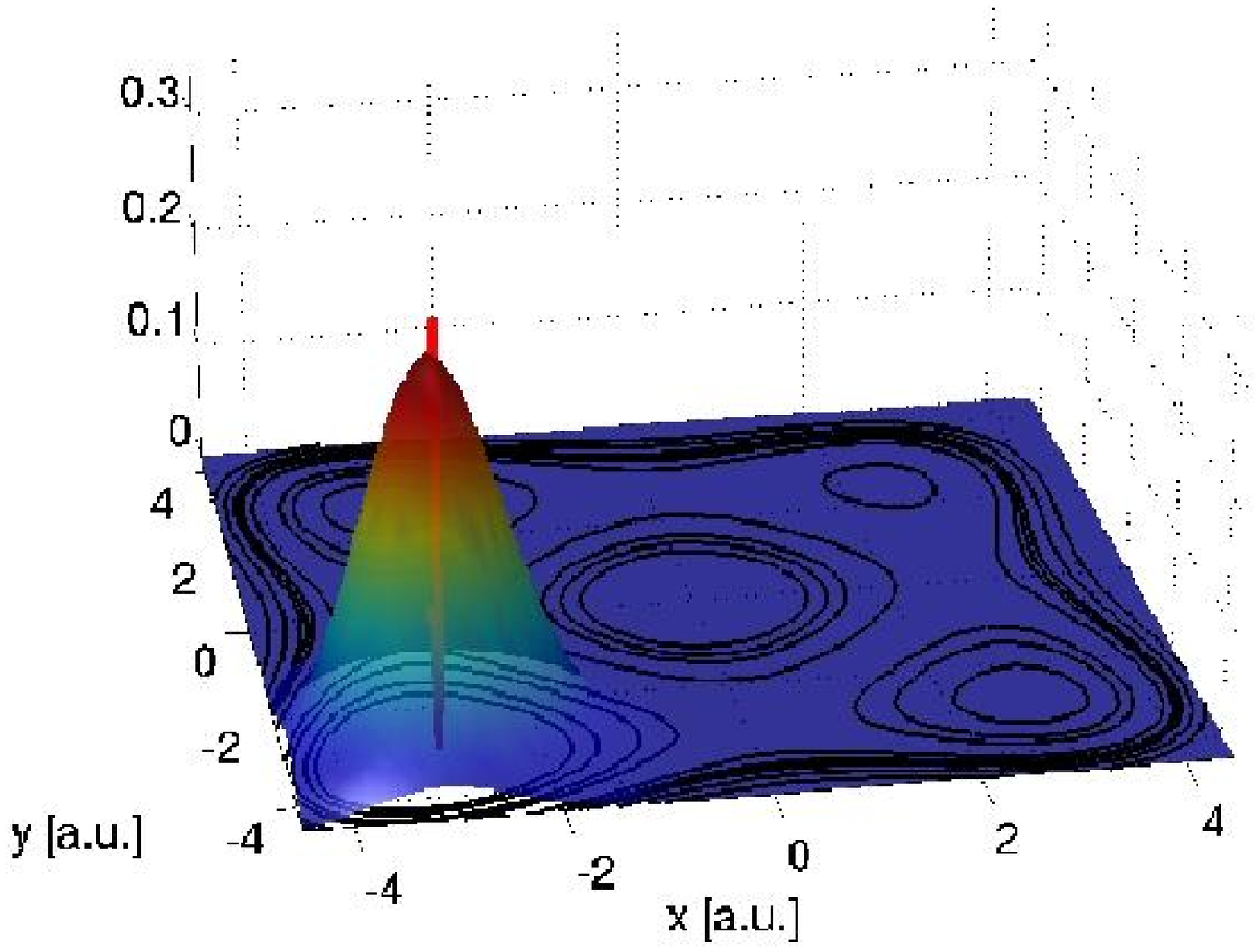}}
}
\subfigure[]{
  \label{fig:field2Di_6p2}
\resizebox{0.46\textwidth}{.23\textheight}{\includegraphics{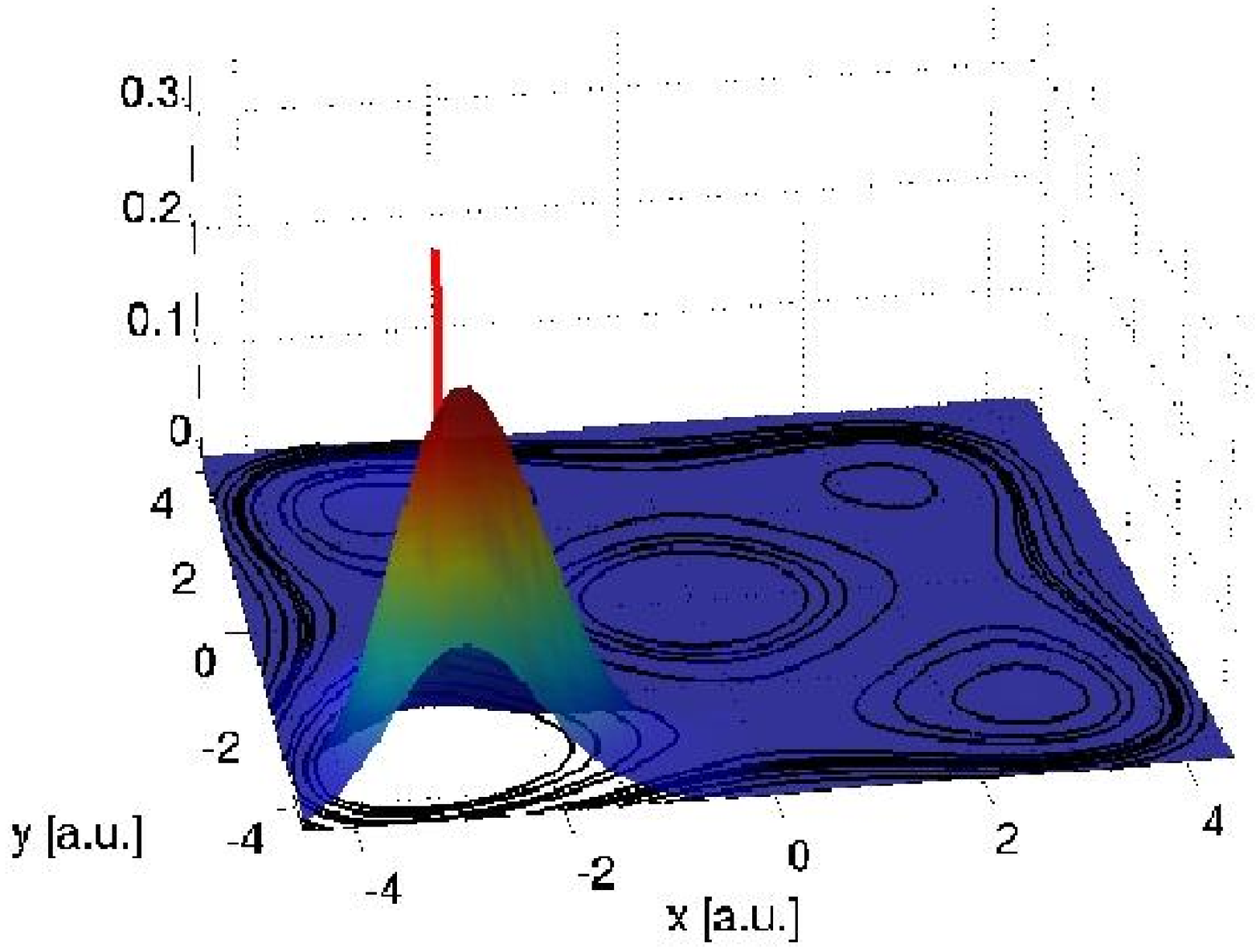}}
}\\
\subfigure[]{
  \label{fig:field2Di_6p3}
 \resizebox{0.46\textwidth}{.23\textheight}{\includegraphics{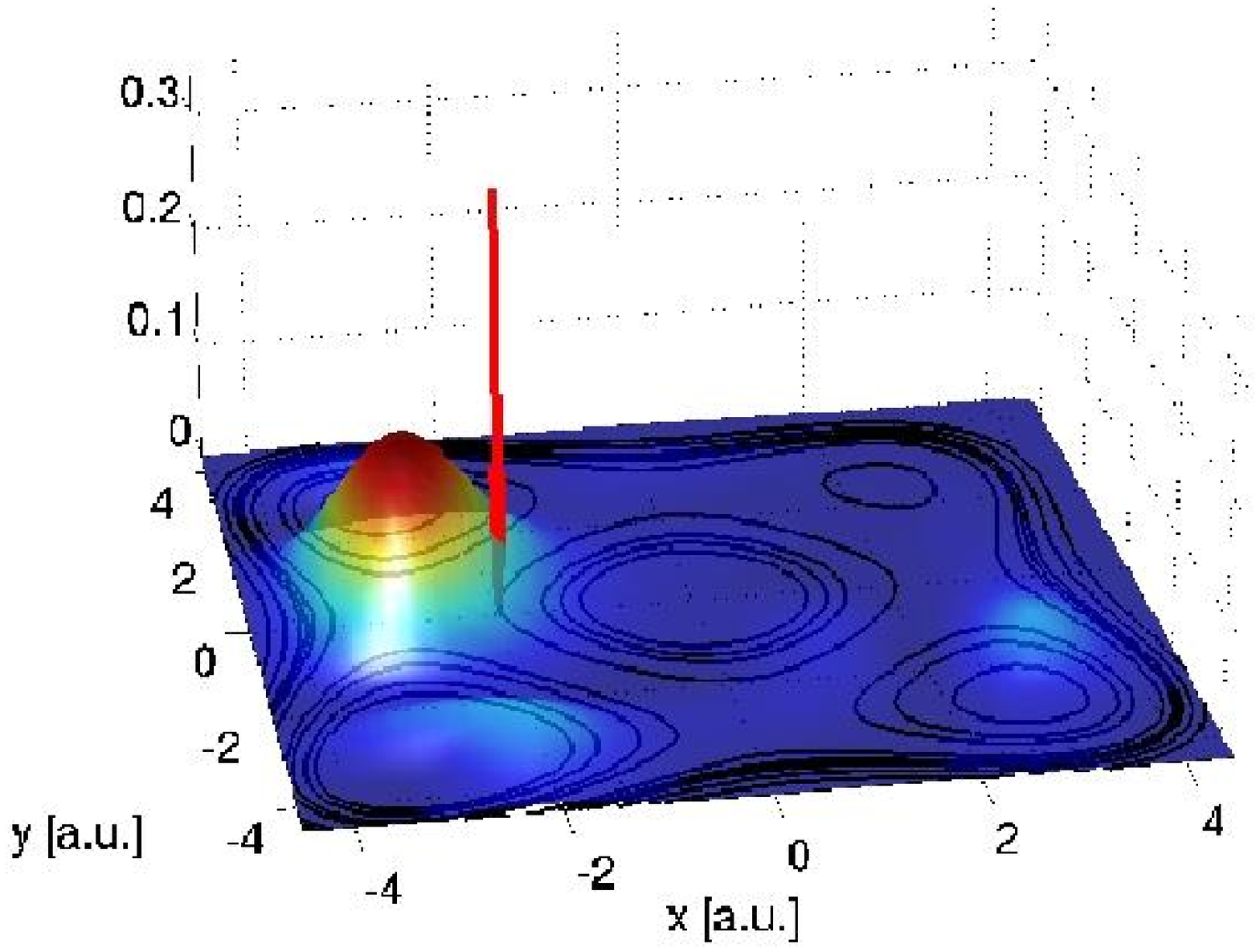}}
}
\subfigure[]{
  \label{fig:field2Di_6p4}
 \resizebox{0.46\textwidth}{.23\textheight}{\includegraphics{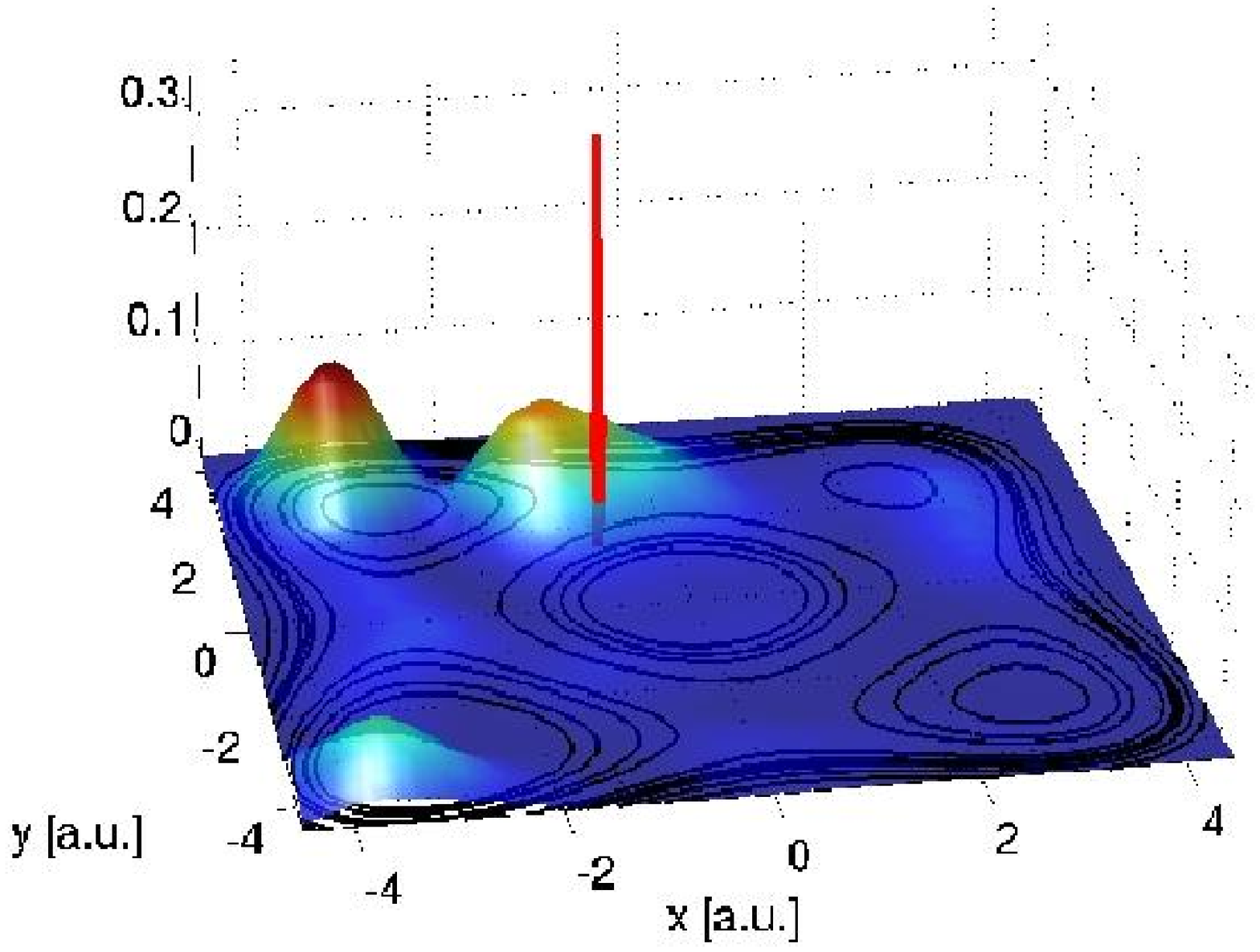}}
}\\
\subfigure[]{
  \label{fig:field2Di_6p5}
 \resizebox{0.46\textwidth}{.23\textheight}{\includegraphics{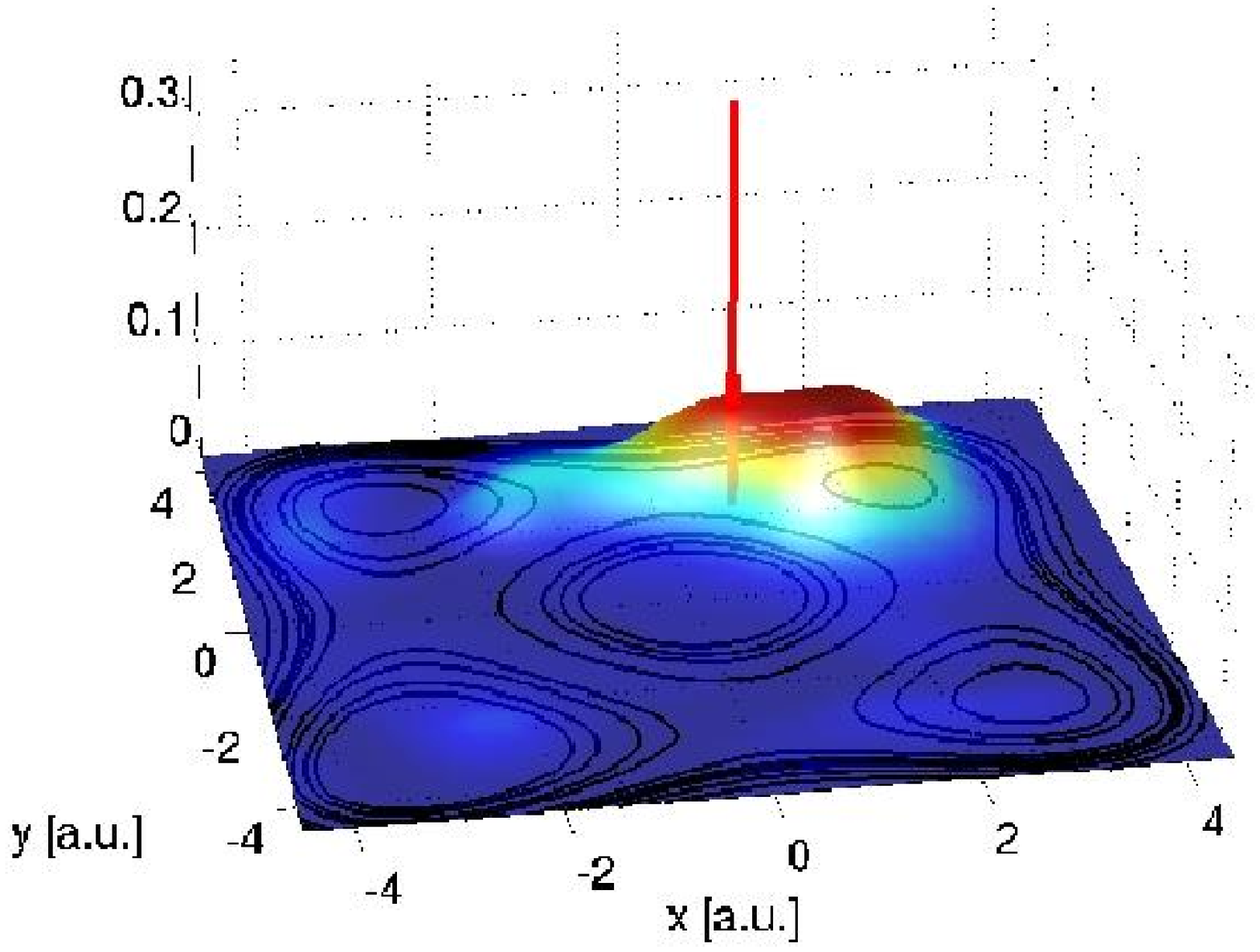}}
}
\subfigure[]{
  \label{fig:field2Di_6p6}
\resizebox{0.46\textwidth}{.23\textheight}{\includegraphics{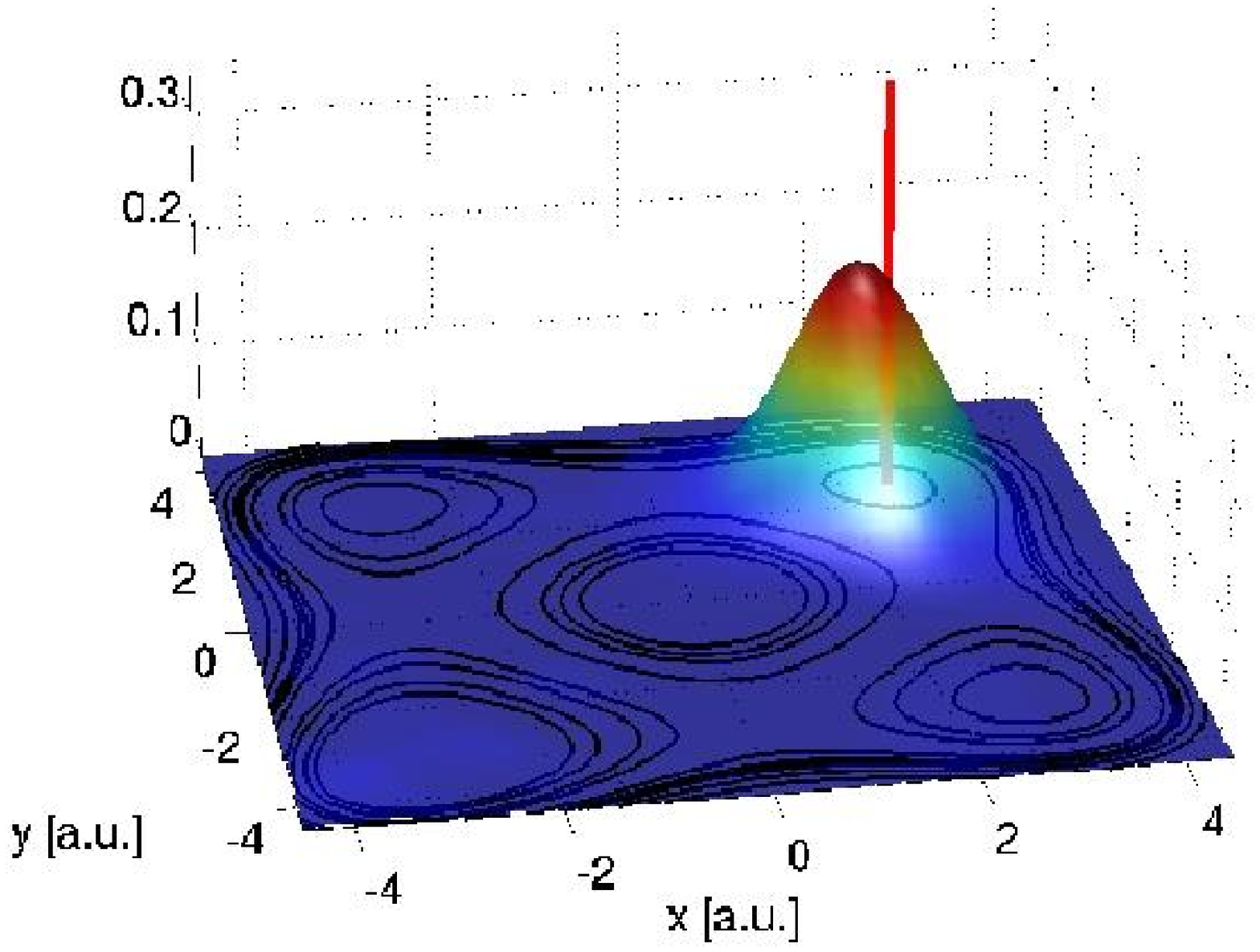}}
}
\caption[Indirect route: Snapshots of the density evolution in time.]{(Color online). Snapshots of the time evolution of the density in the potential well. (a) $t=0$, (b) $t=40$, (c) $t=80$, (d) $t=120$, (e) $t=160$, (f) $t=T=200$. The  contour lines indicate the potential surface while the vertical line marks the center of the target operator [see \eref{eq:2dtargetop}].}
\label{fig:field2Di_6pack}
\end{figure}

The value of the functional after $200$ iterations is $J_1 = 0.0287$ which appears to be small compared to the maximum possible value of $1$. However, one has to be aware that $J_1$ is a measure for the shape of the density (which should be comparable to the target operator) and the position in time. So the control objective is quite demanding. To be able to better assess the quality of the pulse we compare the target trajectory with the expectation value of the position operator and the position of the density maximum, presented in the lower panel of \fref{fig:field2Di_4}. The figure shows that the expectation value meanders around the target curve rather closely. The position of the maximum is also close to the target trajectory but sometimes prefers to be in the minimum at $(-2.7,2.7)$ of the potential. Moreover, we analyze the time behavior of the density by plotting the two-dimensional density distribution for different points in time in \fref{fig:field2Di_6pack}. 
Each figure contains a contour plot of the potential surface, the density distribution and a vertical line which marks the center of the target operator, i.e., the density is supposed to be maximized at this point. These figures illustrate that the laser guides the particle around the barrier in the desired way.

%

\subsection{Transfer over the barrier}
%
The trajectory leading directly across the barrier is given by
\bea
\br^{\mathrm{d}}_0(t) = \left(\begin{matrix} 
  -3.0 + 5.5\, t/T   \cr
  -3.0 + 5.5\, t/T   \cr
\end{matrix} \right),  
\eea
with $0 \leq t \leq T$.

\begin{figure}[!h]
\centering 
\subfigure[]{
  \label{fig:field2Dd_1} 
  \resizebox{0.47\textwidth}{.22\textheight}{\includegraphics*{./2Dd_field.eps}}
    }
\subfigure[]{
  \label{fig:field2Dd_2} 
  \resizebox{0.42\textwidth}{.22\textheight}{\includegraphics*{./2Dd_fft.eps}}
}
\subfigure[]{
  \label{fig:field2Dd_3} 
  \resizebox{0.42\textwidth}{.28\textheight}{\includegraphics*{./2Dd_population2.eps}}
}
\subfigure[]{
  \label{fig:field2Dd_4} 
  \resizebox{0.42\textwidth}{.28\textheight}{\includegraphics*{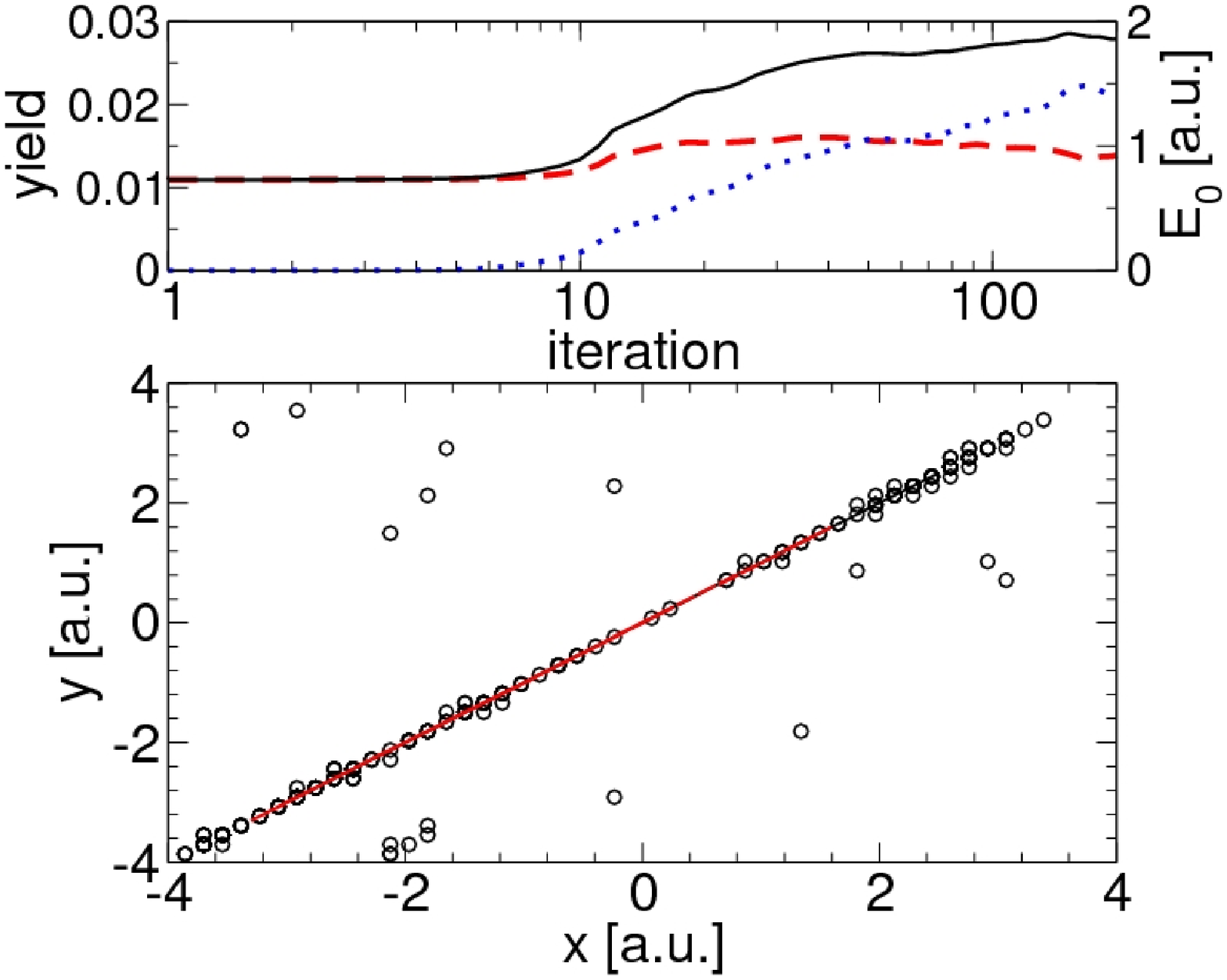}}
}
\caption[Direct route: Results.]{(Color online). Results for the optimization of the direct trajectory [see \fref{fig:pot2Ddirect}]. (a) Optimized laser field. (b) Spectrum of optimized pulse. The $x$ component [(\full) line] and the $y$ component [(\dashed) line] lie on top of each other. (c) Time evolution of the most important occupation numbers: $|\langle n | \Psi(t) \rangle |^2 $: top  panel: $n=0$ (\full), $n= 3$ (\broken), $n=9$ (\opensquare), bottom panel: $n=5$ (\dotted), $n=6$ (\chain), $n=10$ (\opencircle). The (\full) line corresponds to the occupation for all states with $n>15$. (d) Lower panel:  (\dashed) Expectation value of the position operator which lies on top of the target trajectory [(\full) line], (\opencircle) position of the density maximum during the propagation. Upper panel: Convergence of the algorithm. The (\dashed) line corresponds to $J$, the (\full) line to $J_1$, and the (\dotted) line to the laser fluence $E_0$.}
\label{fig:field2Dd}
\end{figure}
The laser pulse shown in \fref{fig:field2Dd_1} is optimized to move the particle directly from the lower left well over the barrier to the upper right well. As one would expect, the $x$ and $y$ components of the laser pulse are identical in that case. 
Again the spectrum, shown in \fref{fig:field2Dd_2}, is difficult to analyze. Compared to the spectrum for the indirect route [\fref{fig:field2Dd_1}] it is broader and therefore even more states are employed in the transition process. This can be seen in \fref{fig:field2Dd_3}, where we plot the occupation numbers for the most important states involved in the transition process. States up to the $20$th excited state have a significant population. In contrast to the previous example the laser needs to excite many delocalized states to achieve a transition across the barrier. 
The convergence shown in the upper panel of \fref{fig:field2Dd_4} is similar to the one found for the indirect route. After 200 iterations we achieve a yield of $J_1=0.0279$ and a fluence of $E_0=1.399$ corresponding to a strong laser field. Again a deviation from the monotonic convergence is observed.

In the lower panel of \fref{fig:field2Dd_4} we plot the expectation value of the position operator [(\dashed) line] and the maximum of the density (\opencircle) both indicate that the transfer of the particle occurs across the barrier. We can also see that a perfect localization cannot always be achieved by the laser, i.e., sometimes the position of the density maximum is found aside of the the target curve. 


\section{SUMMARY} 
\label{sec:conclusion}
We have demonstrated that it is possible to obtain laser pulses optimized to transfer a wavepacket along a predefined path. The laser fields have been calculated with the help of quantum optimal control theory using time-dependent targets.
The pulses in our examples show that a complicated interplay of excitations and de-excitations is necessary to achieve the localization along the given trajectory. Both pulses involve a large number of eigenstates.
Even for this rather simple system an optimization by hand using a combination of $\pi$ pulses\cite{AE75} would be extremely cumbersome while the optimal control method requires only a naive guess for the initial laser field.

In the example where the transfer occurs directly over the barrier the optimization converges to a linearly polarized pulse as one would expect from the symmetry of the problem. The shaping of linearly polarized laser fields is a rather established method. 
However, the optimized pulse for the transfer around the barrier requires a sophisticated time-dependent polarization. The experimental realization of polarization shaped pulses has been demonstrated recently\cite{POS2006, PWWL2006a,PWWL2006b}. Note that technological limitations can also be incorporated in the pulse optimization by combing the method presented above with additional restrictions on the optimal field\cite{WG2005}, like for example spectral constraints\cite{janphd}. 

Currently, we are working on the implementation of the optimal control algorithm into the freely available TDDFT-Solver package \octopus\cite{octopus,octopusurl} which in the future will allow us to investigate the controllability of multi-electron systems as well.

\acknowledgments     
 
This work was supported, in part, by the Deutsche Forschungsgemeinschaft and by the NANOQUANTA Network of Excellence of the European Union.


\bibliography{spieproceedings}   
\bibliographystyle{spiebib}   

\end{document}